\documentclass[aps,prc,twocolumn,superscriptaddress,showpacs]{revtex4}
\usepackage{graphicx}
\usepackage{dcolumn}

\begin{document}

\title{Lifetime measurements of Triaxial Strongly Deformed bands in $^{163}$Tm}

\author{X. Wang}
\affiliation{Physics Department, University of Notre Dame, Notre Dame, IN 46556}
\affiliation{Physics Division, Argonne National Laboratory, Argonne, IL 60439}
\author{R. V.~F.~Janssens}
\author{E. F.~Moore}
\affiliation{Physics Division, Argonne National Laboratory, Argonne, IL 60439}
\author{U. Garg}
\author{Y. Gu}
\author{S. Frauendorf}
\affiliation{Physics Department, University of Notre Dame, Notre Dame, IN 46556}
\author{M. P.~Carpenter}
\affiliation{Physics Division, Argonne National Laboratory, Argonne, IL 60439}
\author{S. S.~Ghugre}
\affiliation{UGC-DAE Consortium for Scientific Research, Kolkata Center, Kolkata 700 098, India}
\author{N. J.~Hammond}
\author{T. Lauritsen}
\affiliation{Physics Division, Argonne National Laboratory, Argonne, IL 60439}
\author{T. Li}
\affiliation{Physics Department, University of Notre Dame, Notre Dame, IN 46556}
\author{G. Mukherjee}
\affiliation{Physics Division, Argonne National Laboratory, Argonne, IL 60439}
\author{N. S.~Pattabiraman}
\affiliation{UGC-DAE Consortium for Scientific Research, Kolkata Center, Kolkata 700 098, India}
\author{D. Seweryniak}
\author{S. Zhu}
\affiliation{Physics Division, Argonne National Laboratory, Argonne, IL 60439}

\date{\today}

\begin{abstract}
With the Doppler Shift Attenuation Method, quadrupole transition moments, 
$Q_t$, were determined for the two recently proposed Triaxial Strongly 
Deformed (TSD) bands in $^{163}$Tm. The measured $Q_t$ moments indicate that 
the deformation of these bands is larger than that of the yrast, signature 
partners. However, the measured values are smaller than those predicted 
by theory. This observation appears to be valid for TSD bands in several 
nuclei of the region. 
\end{abstract}


\maketitle

\section{\label{sec:intro}Introduction}

The collective behavior of nuclei with a triaxial shape has attracted much 
attention over the years. Unfortunately, conclusive experimental signatures 
for such a shape have proven difficult to establish. Nevertheless, pairs of 
nearly degenerate $\Delta I = 1$ bands of the same parity seen in nuclei 
with mass A $\sim$ 100~\cite{Vaman-PRL-92-032501-04,Timar-PRC-73-011301-06,Alcant-PRC-69-024317-04} 
and A $\sim$ 130~\cite{Koike-PRC-67-044319-03,Starosta-PRL-86-971-01,Zhu-PRL-91-132501-03} 
have been interpreted as chiral partner bands, where breaking of the chiral 
symmetry is only possible for a triaxial nuclear shape. Also, rotational 
sequences measured in A $\sim$ 160 -- 175 nuclei~\cite{Bringel-EPJA-16-155-03,%
Odegard-PRL-86-5866-01,Jensen-NPA-703-3-02,Jensen-PRL-89-142503-02,%
Tormanen-PLB-454-8-99,Schonwa-PLB-552-9-03,Amro-PLB-553-197-03,%
Amro-PLB-506-39-01,Neuber-EPJA-15-439-02,Djong-PLB-560-24-03,Hartley-PLB-608-31-05} 
have been associated with the triaxial shapes predicted by theory for 
this mass region. Very recently, it has been 
proposed that the very high spin rotational bands discovered 
beyond band termination in $^{157,158}$Er are also associated with 
the rotation of triaxial shapes~\cite{Paul-PRL-98-012501-07}. 
The strongest experimental evidence for rotation of a triaxial nucleus 
available thus far has been found in a number of Lu isotopes, where 
bands with characteristic properties have been associated with the wobbling 
mode~\cite{Odegard-PRL-86-5866-01,Jensen-NPA-703-3-02,Jensen-PRL-89-142503-02,%
Schonwa-PLB-552-9-03,Amro-PLB-553-197-03}. 
In triaxial nuclei, where different moments of inertia are 
associated with the three principal axes, rotation about the three 
axes is quantum mechanically possible. Although rotation about the axis with the 
largest moment of inertia is favored, the contributions from rotations about 
the other two axes can force the rotation angular momentum vector (R) off the 
principal axis to create a precession or wobbling mode of which the classical 
analog is the rotation of an asymmetric top. As a result, 
a triaxialy deformed nucleus will exhibit a family of rotational 
bands (with identical moments of inertia) based on the same configuration, but 
with different wobbling phonon number ($n_w = 0,1,2,...$)~\cite{Hagemann-NPN-13-20-03}. 
Because the observed sequences also have relatively large deformation, 
they are often referred to as triaxial strongly deformed (TSD) bands.

Until very recently, the fact that wobbling had only been observed in the Lu 
isotopes, and not in any of the neighboring Tm, Ta and Hf nuclei of the region, 
remained somewhat of a puzzle. TSD bands have been reported in many of these 
isotopes~\cite{Amro-PLB-506-39-01,Neuber-EPJA-15-439-02,Djong-PLB-560-24-03,%
Hartley-PLB-608-31-05,Fetea-NPA-690-239-01,Roux-PRC-63-024303-01}, 
but none of them was found to exhibit deexcitation properties 
characteristic of wobbling. In particular, the interband transitions that 
provide a clear signature for wobbling in the Lu isotopes are absent. A 
possible resolution of the issue has recently been proposed in Ref.~\cite{Pattabi-PLB-163Tm-07}, 
following new experimental data on the $^{163}$Tm nucleus. In this work, two 
strongly interacting TSD bands were identified. These sequences were found to 
be linked by several interband transitions. However, these connecting 
$\gamma$ rays do not exhibit properties similar to the ones characteristic 
of wobbling. Rather, they are akin to what would be expected for 
collective structures associated with particle-hole (ph) 
excitations in a TSD well. In Ref.~\cite{Pattabi-PLB-163Tm-07}, this interpretation 
was backed by calculations carried out within the framework 
of the Tilted-Axis Cranking Model. These calculations not only reproduced the 
experimental observables for $^{163}$Tm, but also provided a plausible 
explanation for the presence of wobbling bands in the $_{71}$Lu isotopes and 
their absence in the $_{69}$Tm, $_{72}$Hf and $_{73}$Ta neighbors. Indeed, the 
possibility to identify wobbling bands experimentally is restricted by the 
competition of these collective excitations with the ph excitations. Only if 
the excitation energy above the yrast line of the ph configurations is higher 
than the corresponding energy of the wobbling bands will the latter be fed 
appreciably in fusion-evaporation reactions. This appears to be the case only 
in $_{71}$Lu, where the Fermi surface lies on the i$_{13/2}$ orbital and 
there is a wide gap in the proton level density. 

The purpose of the present paper is two-fold. In Ref.~\cite{Pattabi-PLB-163Tm-07}, 
the interpretation of the two $^{163}$Tm sequences as TSD bands 
rested solely on indirect experimental indications 
(such as the magnitude and evolution with frequency 
of the moments of inertia) and on the agreement with the calculations. Here, 
direct experimental evidence that the two bands are associated with a larger 
deformation than the yrast sequence is provided through the measurement of 
transition quadrupole moments. Furthermore, this work provides an 
additional test of the calculations through a comparison of the calculated 
and measured moments. 

\section{\label{sec:experiment}Experiment}

The experiment was performed at the Argonne 
Tandem Linac Accelerator System (ATLAS). A beam of 165 $MeV$ $^{37}$Cl 
was used to bombard a ``thick'' target, which consisted of a 
0.813 mg/cm$^2$ thick $^{130}$Te layer (isotopic enrichment $\ge$ $95\%$) 
evaporated on a 15 mg/cm$^2$ thick Au foil backed by a 15 mg/cm$^2$ 
layer of Pb. The states of interest in $^{163}$Tm were fed in the 
$^{130}Te(^{37}Cl,4n)$ reaction and their mean lifetimes were 
measured with the Doppler Shift Attenuation Method (DSAM). 
The thickness of the Au backing was chosen such that the evaporation 
residues came to a full stop within this Au layer, while the 
projectiles came to rest in the additional Pb foil. 

In the six-day run, over 1.5 $\times$ 10$^9$ coincidence events with fold $\ge$ 3 
($\it{i.e.}$, with at least three prompt coincident $\gamma$ rays) were collected by 
the Gammasphere detector array~\cite{Janssens-NPN-6-9-96}. Since the DSAM technique 
involves the detection of $\gamma$ rays during the slowing down process 
in the thick target, the relation between the average energy shifts and 
detector angles needs to be determined. For this purpose, the raw data 
was sorted into several BLUE~\cite{Cromaz-NIMA-462-519-01} database files. Unlike 
traditional data-storage techniques for high-fold $\gamma$-ray 
coincidence events, such as the RADWARE software package~\cite{Radford-NIMA-361-306-95}, 
the BLUE database stores the data in its original fold without 
unfolding. Thus, each BLUE file corresponds to the ensemble of all 
coincidence events of a given fold, and each element in the event remains 
encoded not only with the $\gamma$-ray energy and time information, 
but also with the auxiliary information, $\it{e.g.}$, the detector 
identification. The specific data structure of BLUE is such that 
producing background-subtracted spectra at a given detector angle under 
specific coincidence requirements can be achieved efficiently with the 
method described in Ref.~\cite{Starosta-NIMA-515-771-03}. 

\section{\label{sec:results}Data Analysis and Results}

\begin{figure*}
\begin{center}
\includegraphics[angle=270,width=1.50\columnwidth]{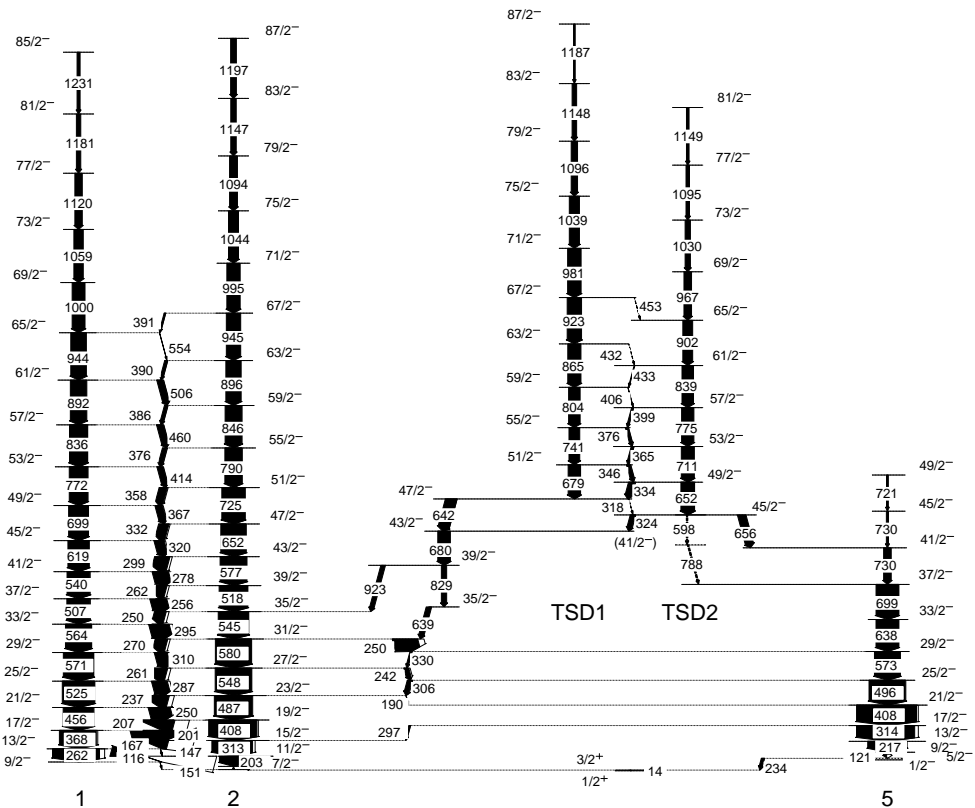}
\caption{Partial level scheme of $^{163}$Tm. Adapted from 
Ref.~\cite{Pattabi-PLB-163Tm-07}.\label{fig:163Tm_level_scheme_PRC}}
\end{center}
\end{figure*}

\begin{figure}
\includegraphics[angle=0,width=\columnwidth]{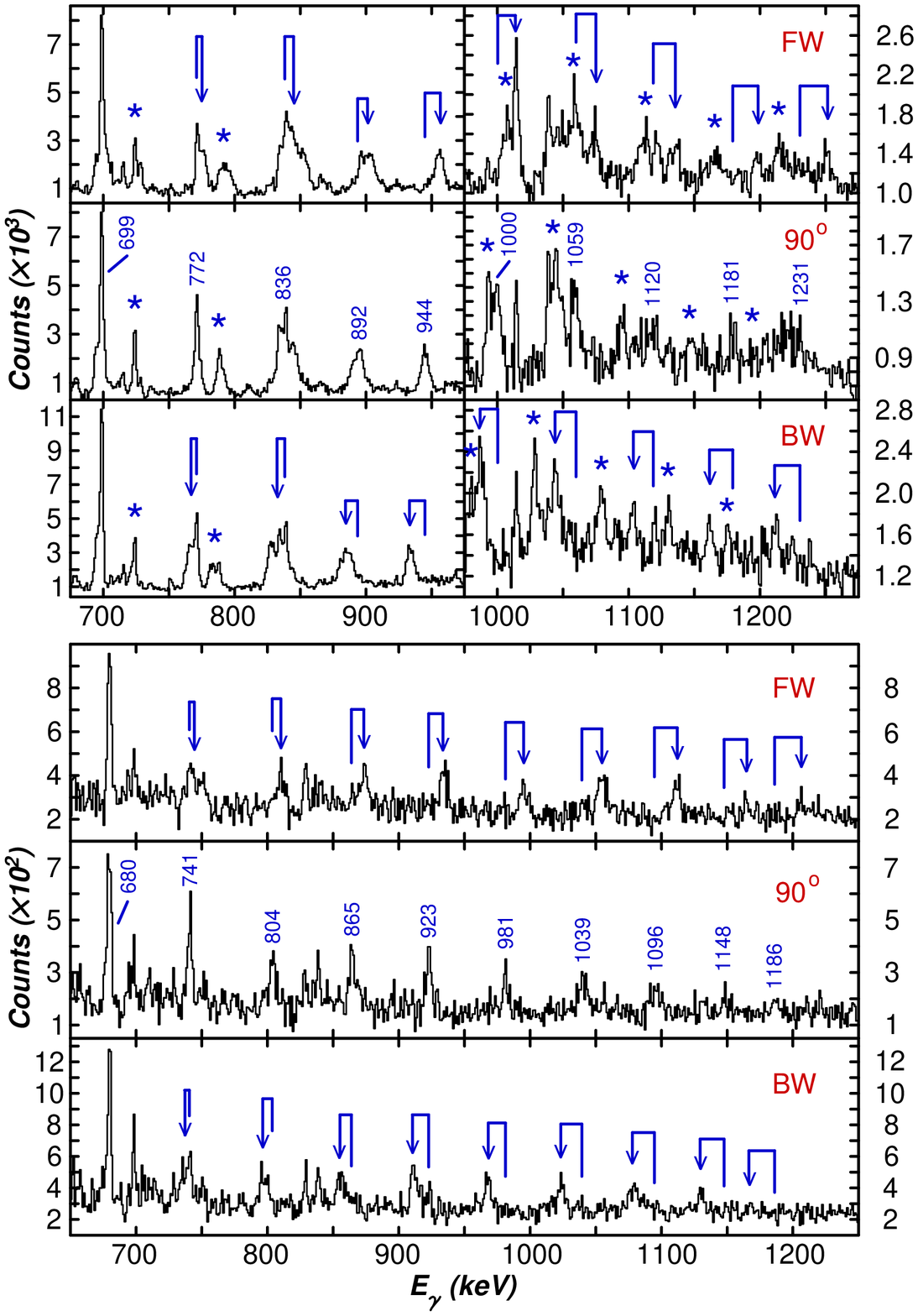}
\caption{(Color on-line) Sum of spectra gated on in-band transitions for 
bands 1 (top) and TSD1 (bottom) at 3 detector angles: 35$^{\circ}$ 
(FW), 90$^{\circ}$, and 145$^{\circ}$ (BW). The positions of unshifted and 
shifted $\gamma$ rays are marked by energy values and arrows, respectively. 
Note that transitions from band 2 appear in the band 1 spectra (marked with 
$\star$ symbols) due to the fact that intense connecting transitions occur 
between the two bands, as reported in 
Ref.~\cite{Pattabi-PLB-163Tm-07}.\label{fig:DSAMspectra}}
\end{figure}

The present work focuses on four rotational bands delineated in 
Ref.~\cite{Pattabi-PLB-163Tm-07}, and shown in 
Fig.~\ref{fig:163Tm_level_scheme_PRC}. Hence, the nomenclature proposed in that 
paper has been adopted here, $\it{i.e.}$, the two bands associated 
with the $[523]7/2^{-}$ configuration of normal deformation 
are designated as band 1 ($85/2^{-}$ -- $9/2^{-}$ sequence) 
and 2 ($87/2^{-}$ -- $7/2^{-}$ cascade), while the proposed 
triaxial bands are labeled as TSD1 
($87/2^{-}$ -- $47/2^{-}$ sequence) and TSD2 
($81/2^{-}$ -- $45/2^{-}$ cascade), respectively. From a 
first inspection of the coincidence data, it was established 
that the transitions with energy $E_{\gamma}$ $\le$ 600 $keV$ in 
bands 1 and 2 did not exhibit any measurable shift or broadening 
as a function of the detector angle. In other words, these 
deexcitations must have occurred after the recoiling nuclei 
have come to rest in the Au layer of the target. These ``stopped'' 
transitions could thus be used as a starting point to obtain 
coincidence spectra for each band at ten detector angles, from 
which energy shifts would be determined. The use of "stopped" 
transitions alone proved to be insufficient. Hence, angle-dependent 
gates had to be placed on band members in an iterative procedure 
starting with the lowest $\gamma$ ray exhibiting a shift and 
moving up in the band one transition at each step. This procedure 
could be applied not only to bands 1 and 2, but also the TSD1 and 
TSD2 sequences since the latter deexcite into bands 1 and 2. In the 
process of selecting appropriate gating conditions, special care 
was taken to avoid numerous contaminant lines from either 
other $^{163}$Tm band structures or other reaction products, as well 
as some in-band doublet $\gamma$ rays such as the 680-$keV$ line in 
TSD1, for example, which corresponds both to the 
$51/2^{-}$$\rightarrow$$47/2^{-}$ transition and to the 
$43/2^{-}$$\rightarrow$$39/2^{-}$ transition 
in the decay sequence toward bands 1 and 2~\cite{Pattabi-PLB-163Tm-07}. In this 
context, the analysis of band TSD2 proved to be particularly challenging 
as it is the one most affected by the closeness in energy of many 
in-band transitions with either those in band 2 or other contaminant 
peaks. Proceeding in this careful manner, an optimized spectrum 
was obtained at each detector angle by summing up all clean 
double-gated coincidence spectra with the appropriate gating 
conditions. Representative spectra resulting from this analysis are 
presented in Fig.~\ref{fig:DSAMspectra} for bands 1 and TSD1. 

\begin{figure}
\includegraphics[angle=270,width=\columnwidth]{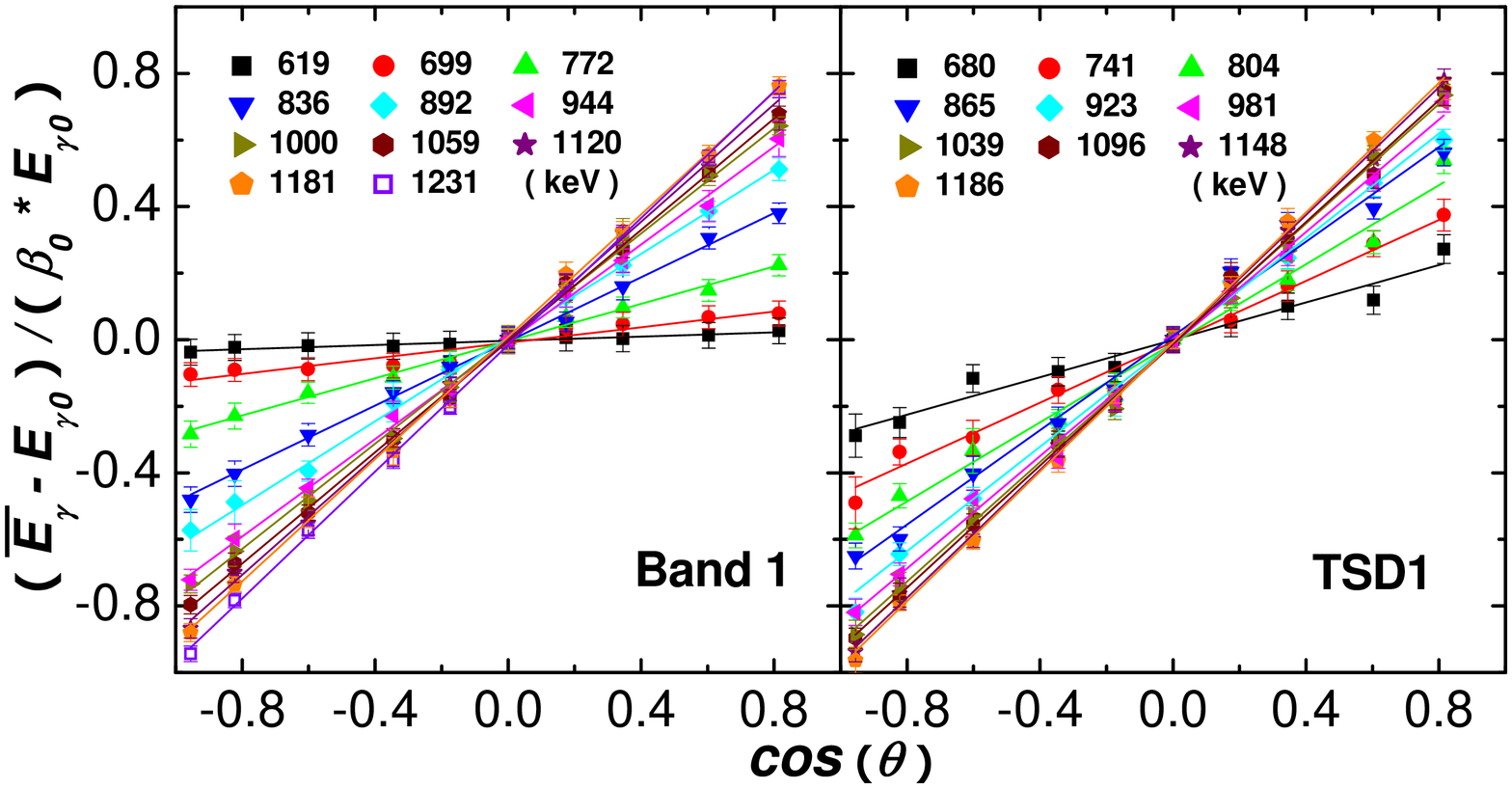}
\caption{(Color on-line) Linear fits to the $\gamma$-ray energy shifts 
as a function of $cos(\theta)$ for bands 1 and TSD1.\label{fig:linear}} 
\end{figure}

\begin{figure}
\includegraphics[angle=270,width=\columnwidth]{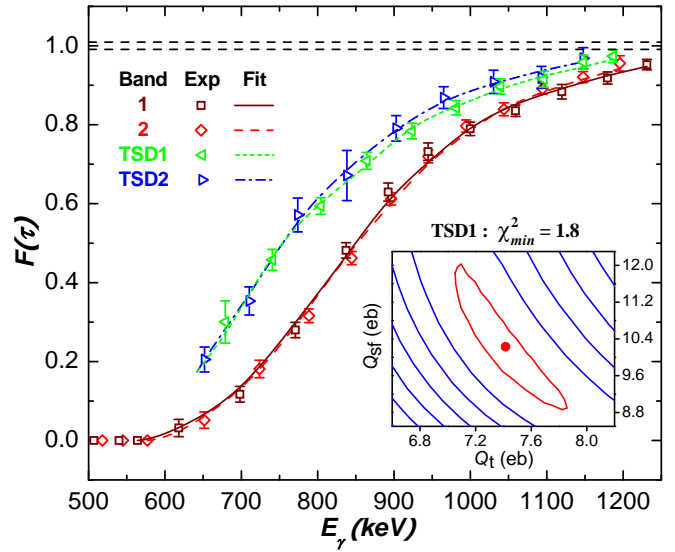}
\caption{(Color on-line) Measured $F(\tau)$ values with best-fit curves 
as described in the text for the four bands in $^{163}$Tm. The two 
horizontal dashed lines show the range of initial recoil velocities within 
the $^{130}$Te target layer. Insert: plot of the ${\chi}^2(Q_t,Q_{sf})$ 
surface for band TSD1. The central dot indicates the location of the minimum 
$({\chi}^{2}_{min}=1.8)$, with the first contour plotted in an increment 
of one.\label{fig:curve}}
\end{figure}

The fractions of full Doppler shift $F(\tau)$ and the associated errors 
were subsequently extracted for transitions in the four $^{163}$Tm bands 
of interest through linear fits of the shifts measured at 10 angles with 
the expression
\[
F(\tau) = \frac{\overline{E_\gamma} - E_{\gamma 0}}{E_{\gamma 0} \ast \beta_0 \ast \cos(\theta)}
\]~\cite{Moore-ZPA-358-219-97}. 
Here, for every transition $E_{\gamma 0}$ is the nominal $\gamma$-ray energy, 
$\overline{E_\gamma}$ is the measured energy at the angle $\theta$, and 
$\beta_0$ is the initial recoil velocity of the $^{163}$Tm residues formed 
in the center of the $^{130}$Te target layer. This quantity was calculated 
to be ${\beta_0}={v_0 / c}=0.02148$ with the help of the stopping powers 
computed with the code SRIM 2003~\cite{Ziegler-85-book}. Samples of the linear fits 
can be seen in Fig.~\ref{fig:linear} for bands 1 and TSD1. The resulting 
$F(\tau)$ values are presented as a function of the $\gamma$-ray energy for 
all four bands in Fig.~\ref{fig:curve}. A cursory inspection of this figure 
indicates two families of $F(\tau)$ curves: for similar transition 
energies bands 1 and 2 have distinctly smaller values than bands TSD1 and 
TSD2. It is also worth noting that the larger $F(\tau)$ uncertainties 
associated with band TSD2 relate to the difficulty of obtaining 
suitable spectra as discussed above. 

The intrinsic transition quadrupole moments $Q_t$ of the four bands were 
extracted from the measured $F(\tau)$ values using the new Monte Carlo computer 
code MLIFETIME. As is usually the case in this type of analysis, a number of 
assumptions were made in order to compute the average recoil velocity at which 
the decay from a particular state occurs. These are: (1) all levels in a given 
band have the same transition quadrupole moment $Q_t$; (2) the sidefeeding 
into each level in a band is modeled as a single cascade with a common, 
constant quadrupole moment $Q_{sf}$, and characterized by the same dynamic 
moment of inertia $\Im^{(2)}$ as the main band into which they feed; the 
number of transitions in each sidefeeding band is proportional to the number 
of transitions in the main band above the state of interest; (3) the sidefeeding 
intensities are determined directly from the measured $\gamma$-ray intensities 
within the bands; and (4) a one-step delay at the top of all feeder bands was 
parameterized by a single lifetime $T_{sf}$. 

The detailed slowing-down histories of the recoiling $^{163}$Tm ions in both 
the target and the Au backing were calculated using the SRIM 2003~\cite{Ziegler-85-book} 
Monte Carlo code. The initial positions and velocity vectors for each of 10,000 
starting ions were calculated in a Monte Carlo fashion which included the 
broadening of the recoil cone due to the evaporation of neutrons from the 
$^{167}$Tm compound nucleus. The production cross section was assumed to be 
constant over the range of energies due to the beam slowing down in the target. 
This resulted in an even distribution for the starting positions of the 
$^{163}$Tm ions throughout the target thickness. The initial $^{163}$Tm ion 
positions in the target, ion energies, and recoil direction were supplied as 
input to the SRIM 2003 code, which then transported each ion through the 
target and the backing. The detailed recoil history for each ion was 
written out to a computer file which listed the energy, direction, and position 
at which each collision of the recoiling $^{163}$Tm ions with the target and 
backing atoms occurred. The lifetime code then read in this file and tracked 
each ion history in one femto-second (1 $fs$) time steps from initial 
formation until the ion came to rest. In order to compute the Doppler shifted 
energies of each $\gamma$ ray emitted by the recoiling $^{163}$Tm ions in a 
Monte-Carlo fashion, feeder bands into each state in the main band were randomly 
populated according to the measured intensity distribution. The subsequent 
decay profile through the feeder and main band was tracked in 1 $fs$ steps, 
with the decay probability given by the radioactive decay law using the 
$T_{sf}$ parameter and lifetimes generated from each $Q_{sf}$, $Q_t$ parameter 
set. The velocity vector of the $\gamma$ emitting ion was recorded at the 
time of decay of each state of interest. The calculated average fraction of 
the full Doppler shift was generated by accumulating a large number of histories. 
In the present analysis, each of the 10,000 ion histories was used 10 times, 
resulting in better than $1\%$ statistical uncertainty in the 
calculated $F(\tau)$ values. 

A ${\chi}^2$ minimization using the fit parameters 
$Q_t$, $Q_{sf}$ and $T_{sf}$ was performed to the measured $F(\tau)$ values 
for the four bands. The results of the fitting process are summarized 
in Table~\ref{tab:fitmoments}, where the quoted errors include the 
covariance between the fit parameters. As can be seen from Fig.~\ref{fig:curve}, 
the fit of the $F(\tau)$ data is satisfactory in all cases. This is illustrated 
further in the case of band TSD1 in the insert to Fig.~\ref{fig:curve}, where 
contours of ${\chi}^2$ values are presented in a $(Q_t,Q_{sf})$ plane and 
a clear minimum can be seen.

\begin{table}
\caption{Summary of quadrupole moments resulting from DSAM centroid 
shift analysis for the 4 bands in $^{163}$Tm. In all cases the value 
of $T_{sf}$ is very small, $\it{i.e.}$, $T_{sf}~{\sim}~1~fs$. The error bars are 
statistical only, $\it{i.e.}$, they do not include the $\sim$ $15\%$ error associated with 
the systematic uncertainty in the stopping powers (see text for details).\label{tab:fitmoments}}
\begin{ruledtabular}
\begin{tabular}{cccc}
Band & $Q_t~(eb)$ & $Q_{sf}~(eb)$ & ${\chi}^{2}_{min}$\\ \hline
\\
1 & $6.40^{+0.57}_{-0.33}$ & $6.74^{+0.73}_{-0.84}$ & 6.51\\
\\
2 & $6.39^{+0.33}_{-0.31}$ & $6.97^{+0.91}_{-0.63}$ & 8.01\\
\\
TSD1 & $7.42^{+0.44}_{-0.37}$ & $10.23^{+1.79}_{-1.34}$ & 1.81\\
\\
TSD2 & $7.70^{+1.04}_{-0.57}$ & $9.65^{+2.85}_{-2.25}$ & 1.15\\
\end{tabular}
\end{ruledtabular}
\end{table}

\section{\label{sec:discussion}Discussion}

Before discussing the significance of the difference in the measured $Q_t$ values for 
bands 1 and 2, on the one hand, and the TSD1 and TSD2 sequences on the other, it is worth 
examining the relevance of the results through a comparison with other nuclei in the 
region. Since bands 1 and 2 are based on the $[523]7/2^{-}$ configuration, a search of 
the literature was undertaken for quadrupole moment measurements of this configuration 
in neighboring nuclei. The results are given in Table~\ref{tab:NDcompare}. The 
$[523]7/2^{-}$ configuration is yrast in $^{163,165}$Ho, and, perhaps more importantly, 
in $^{163}$Lu, one of the isotopes where TSD and wobbling bands are known as well. It 
should be noted that the $^{163}$Lu yrast sequence had first been associated with the 
$[514]9/2^{-}$ configuration~\cite{Schmitz-NPA-539-112-92}. However, following the work of 
Ref.~\cite{Schmitz-PLB-303-230-93}, the $[523]7/2^{-}$ configuration was adopted on the basis of 
the $B(E2)$ and $B(M1)$ transition probabilities deduced from the measured lifetimes 
and branching ratios. 

\begin{table}
\caption{Quadrupole moments of ND bands based on the $[523]7/2^{-}$ configuration 
in Tm, Ho, and Lu nuclei. The last column provides the reference and identifies the 
method used to measure the moments by the following symbols: FT - DSAM $F(\tau)$; 
LS - DSAM line shape; RD - Recoil distance; LRIMS - Laser resonance ionization; 
KaX - Kaonic X-ray; PiX - Pionic X-ray; MuX - Muonic X-ray. The error bars 
are statistical only and do not include the systematic uncertainty in the 
stopping powers. Note that for some entries in the table, a range of values is given. 
The reader is referred to the cited reference for further details.\label{tab:NDcompare}}
\begin{ruledtabular}
\begin{tabular}{cccc}
Nuclide & Band & $Q_t~(eb)$ & Method~[REF]\\ \hline
\\
$^{163}$Tm & 1 & $6.40^{+0.57}_{-0.33}$ & FT~[present work]\\
\\
$^{163}$Tm & 2 & $6.39^{+0.33}_{-0.31}$ & FT~[present work]\\
\\
$^{163}$Ho & ND & $6.78\pm1.13$ & LRIMS~\cite{Alkh-NPA-504-549-89}\\
\\
$^{165}$Ho & ND1 & $6.42\pm0.15$, $6.78\pm0.04$ & KaX, PiX~\cite{Batty-NPA-355-383-81}\\
\\
 & & $6.74\pm0.04$ & PiX~\cite{Olani-NPA-403-572-83}\\
\\
 & & $6.57\pm0.06$ & MuX~\cite{Powers-NPA-262-493-76}\\
\\
$^{165}$Ho & ND2 & $5.76\pm0.07$ & MuX~\cite{Powers-NPA-262-493-76}\\
\\
$^{163}$Lu & ND1 & $4.88^{+1.36}_{-0.68}$ -- $6.78^{+2.66}_{-1.39}$ & LS + RD~\cite{Schmitz-PLB-303-230-93}\\
\\
$^{163}$Lu & ND2 & $2.13^{+0.62}_{-0.43}$ -- $6.72^{+0.77}_{-0.40}$ & LS + RD~\cite{Schmitz-PLB-303-230-93}\\
\end{tabular}
\end{ruledtabular}
\end{table}

As can be seen from Table~\ref{tab:NDcompare}, the $Q_t$ moments have been obtained 
using a number of techniques ranging from the analysis of $F(\tau)$ values, such as 
those in the present work, and full line shape analyses of data taken using the 
DSAM technique, to measurements with the recoil distance method ($\it{e.g.}$, so-called 
plunger data), and even to laser resonance ionization as well as detection of the 
characteristic $X$ rays of kaonic, pionic or muonic atoms. It can be concluded from 
Table~\ref{tab:NDcompare} that the moments measured in the present work for bands 
1 and 2 ($Q_t$ $\sim$ $6.4~eb$) are in good agreement with those reported for the same 
configuration in the literature. This observation provides further confidence in the 
analysis presented above. 

The $Q_t$ moments of bands 1 and 2 can then also be compared 
with the calculations first outlined in Ref.~\cite{Pattabi-PLB-163Tm-07}. These predict the value 
to be $Q_t$ = $5.8~eb$ at spin $I$ = 30, with an associated axial quadrupole deformation 
of ${\epsilon}_2$ = 0.21. Considering the fact that the errors quoted for the 
$Q_t$ moments in Table~\ref{tab:fitmoments} are statistical only and do not include 
the additional systematic error of $\sim$ $15\%$ due to the uncertainties in the stopping 
powers, the agreement between experiment and theory can be considered as satisfactory. 
Nevertheless, the fact remains that deformations calculated with the Cranked Nilsson-Strutinsky 
(CNS) model~\cite{Bengtsson-NPA-436-14-85}, the Tilted-Axis Cranking (TAC) model~\cite{Frauendorf-NPA-677-115-00} 
or the Ultimate Cranker (UC) code~\cite{Bengtsson-UC-code-URL}, all using the same Nilsson potential, 
tend to be systematically somewhat smaller than the values derived from experiment, 
an observation that warrants further theoretical investigation. 

The present data clearly indicate that the deformation 
associated with bands TSD1 and TSD2 is larger 
than that of the yrast structure: as can be seen from Table~\ref{tab:fitmoments}, 
the $Q_t$ moments of bands TSD1 and TSD2 ($\sim$ $7.5~eb$) exceed those for bands 
1 and 2 by $\sim$ $1~eb$. The larger deformation agrees with the interpretation 
proposed in Ref.~\cite{Pattabi-PLB-163Tm-07}. However, the magnitude of the increase in $Q_t$ 
moments is not reproduced as the TAC calculations indicate a transitional 
quadrupole moment increasing slightly from $8.7~eb$ at spin $I$ = 24 to $9.6~eb$ 
for 34 $<$ $I$ $<$ 50. At present, this discrepancy between data and calculations 
is not understood. It is, however, not unique to $^{163}$Tm. Table~\ref{tab:TSDcompare} 
compares $Q_t$ moments for TSD bands in all nuclei of the region where this information 
is available. Just as was the case above, the systematic uncertainty associated with 
the stopping powers has been ignored. Nevertheless, three rather striking observations 
can be made from Table~\ref{tab:TSDcompare}: (1) the $Q_t$ values for the TSD bands in 
$^{163}$Lu and $^{163}$Tm are essentially the same, (2) the $Q_t$ moments of the TSD 
bands decrease from $^{163}$Lu and $^{163}$Tm to $^{165}$Lu, an observation already 
made for Lu isotopes in Refs.~\cite{Schonwa-EPJA-13-291-02,Schonwa-EPJA-15-435-02}, and (3) all the TSD bands in 
Hf nuclei are characterized by $Q_t$ moments that are larger than those in Lu and Tm 
by $\sim$ 4 -- $6~eb$, possibly pointing to a rather different nature for these bands. 
Just as in the present $^{163}$Tm case, a discrepancy between the measured and 
calculated $Q_t$ moments was found for the Lu isotopes: UC calculations predicted 
values of $Q_t$ $\sim$ $9.2~eb$ and $11.5~eb$ for positive and negative values of the 
deformation parameter $\gamma$, and these values were computed to be essentially 
the same for the three Lu isotopes ($A$ = 163, 164, 165)~\cite{Schonwa-EPJA-13-291-02,Schonwa-EPJA-15-435-02}, 
but with the configuration associated with a rotation about the short axis 
($\gamma$ $>$ 0) being lower in energy. As stated above, the physical 
origin of the discrepancy between theory and experiment is at present 
unclear, although it was pointed out in Refs.~\cite{Schonwa-EPJA-13-291-02,Schonwa-EPJA-15-435-02} that the 
exact location in energy of the $i_{13/2}$ and $h_{9/2}$ proton- and $i_{11/2}$ 
neutron-intruder orbitals is crucial for the deformation. These orbitals are 
deformation driving and, hence, might have a considerable impact on the $Q_t$ 
moments. It is possible that the use of the standard Nilsson potential parameters, 
questioned above for normal deformed configurations, needs also to be reconsidered 
for the precise description of TSD bands. 

\begin{table*}
\caption{Quadrupole moments of TSD bands in Tm, Lu, and Hf nuclei. 
The last column provides the reference and identifies the method 
used to measure the moments by the following symbols: 
FT - DSAM $F(\tau)$; LS - DSAM line shape. The error bars 
are statistical only and do not include the systematic uncertainty 
in the stopping powers. Note that for some entries in the table, 
a range of values is given. The reader is referred to the cited 
reference for further details.\label{tab:TSDcompare}}
\begin{ruledtabular}
\begin{tabular}{ccccc}
Nuclide & Band & $Q_t~(eb)$ & $Q_{sf}~(eb)$ & Method~[REF]\\ \hline
\\
$^{163}$Tm & TSD1 & $7.42^{+0.44}_{-0.37}$ & $10.23^{+1.79}_{-1.34}$ & FT~[present work]\\
\\
$^{163}$Tm & TSD2 & $7.70^{+1.04}_{-0.57}$ & $9.65^{+2.85}_{-2.25}$ & FT~[present work]\\
\\
$^{163}$Lu & TSD1 & $7.4^{+0.7}_{-0.4}$, $7.7^{+2.3}_{-1.3}$ 
& $6.7^{+0.7}_{-0.7}$, $7.0^{+0.7}_{-0.7}$ & FT~\cite{Schonwa-EPJA-15-435-02}\\
\\
 & & $7.63^{+1.46}_{-0.88}$ -- $9.93^{+1.14}_{-0.99}$ & & LS~\cite{Gorgen-PRC-69-031301-04}\\
\\
$^{163}$Lu & TSD2 & $6.68^{+1.70}_{-1.02}$ -- $8.51^{+0.95}_{-0.73}$ & & LS~\cite{Gorgen-PRC-69-031301-04}\\
\\
$^{164}$Lu & TSD1 & $7.4^{+2.5}_{-1.3}$ & $6.7^{+0.7}_{-0.7}$ & FT~\cite{Schonwa-EPJA-15-435-02}\\
\\
$^{165}$Lu & TSD1 & $6.0^{+1.2}_{-0.2}$, $6.4^{+1.9}_{-0.7}$ 
& $5.4^{+0.5}_{-0.5}$, $5.8^{+0.6}_{-0.6}$ & FT~\cite{Schonwa-EPJA-15-435-02}\\
\\
$^{167}$Lu & TSD1 & $6.9^{+0.3}_{-0.3}$ & $4.4^{+0.4}_{-0.2}$ & FT~\cite{Gurdal-JPG-31-S1873-05}\\
\\
$^{168}$Hf & TSD1 & $11.4^{+1.1}_{-1.2}$ & $10.5^{+1.7}_{-1.6}$ & FT~\cite{Amro-PLB-506-39-01}\\
\\
$^{174}$Hf & TSD1 & $13.8^{+0.3}_{-0.4}$ & $8.4^{+0.3}_{-0.3}$ & FT~\cite{Hartley-PLB-608-31-05}\\
\\
$^{174}$Hf & TSD2 & $13.5^{+0.2}_{-0.3}$ & $8.0^{+0.3}_{-0.2}$ & FT~\cite{Hartley-PLB-608-31-05}\\
\\
$^{174}$Hf & TSD3 & $13.0^{+0.8}_{-0.4}$ & $10.3^{+0.6}_{-0.8}$ & FT~\cite{Hartley-PLB-608-31-05}\\
\\
$^{174}$Hf & TSD4 & $12.6^{+0.8}_{-0.8}$ & $10.2^{+1.6}_{-1.3}$ & FT~\cite{Hartley-PLB-608-31-05}\\
\end{tabular}
\end{ruledtabular}
\end{table*}

In Ref.~\cite{Schonwa-EPJA-13-291-02} it was argued that the fact that the measured $Q_t$ moments 
in the $^{163}$Lu TSD bands are smaller than the calculated values points towards 
a positive $\gamma$ deformation because the latter is associated with the smaller 
computed moments. As already discussed in Ref.~\cite{Pattabi-PLB-163Tm-07}, the same conclusion 
can not be drawn in the case of $^{163}$Tm. Indeed, TAC calculations, which 
do not restrict the orientation of rotational axis to one of the principal axes, 
point to a tilted solution that smoothly connects two minima of opposite sign in 
$\gamma$ deformation. The average deformation parameters are ${\epsilon}_2$ = 0.39, 
$|\gamma|$ = 17$^{\circ}$. For $I$ $>$ 23 the angular 
momentum vector gradually moves away from the intermediate axis ($\gamma$ $<$ 0) 
toward the short one ($\gamma$ $>$ 0), without reaching the latter by $I$ = 50. 

\begin{figure}
\includegraphics[angle=0,width=\columnwidth]{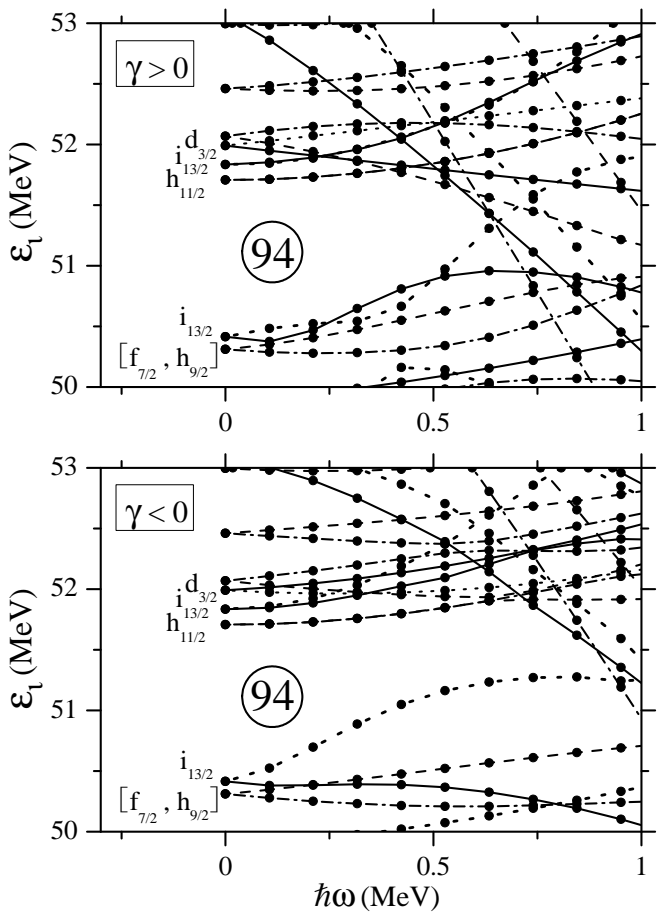}
\caption{Single-neutron routhians as function of 
rotational frequency in TSD minima 1 (top) and 2 (bottom). The line convention is: 
$(\pi,\alpha)=$ (+,1/2) full, (+,-1/2) dot, (-,1/2) dash, (-,-1/2) 
dash dot. The deformation parameters used in the calculations are: 
$\epsilon_{2}= 0.39$, $\epsilon_{4} = 0.05$, $|\gamma|= 17^{\circ}$.\label{fig:163Tm_neutron_routhian}}
\end{figure}

The calculations of Ref.~\cite{Pattabi-PLB-163Tm-07} have also 
been extended to the case of $^{163}$Lu and the computed $Q_t$ moments for the TSD 
bands are larger than the measured ones, in agreement with the general findings 
discussed above. These $Q_t$ moments in $^{163}$Lu were also found to decrease slightly 
from $10.3~eb$ at $I$ = 20 to $9.7~eb$ at $I$ = 40 just as in $^{163}$Tm. Moreover, 
the $^{163}$Lu values are also somewhat larger than the corresponding ones in $^{163}$Tm, 
reflecting the additional drive towards larger deformation brought about 
by the $i_{13/2}$ proton orbital which is occupied in this case. However, it 
should be pointed out that within the framework of these 
calculations~\cite{Pattabi-PLB-163Tm-07}, the occupation of the $i_{13/2}$ 
proton orbital is not a necessary condition to achieve a TSD minimum. Rather, 
the deformation is driven mainly by the $N$ = 94 neutron gap. This point is 
illustrated further in Fig.~\ref{fig:163Tm_neutron_routhian} where the single-neutron 
routhians are presented and the large $N=94$ gap associated with the TSD 
shapes at positive and negative $\gamma$ values is clearly visible. The 
corresponding single-proton routhians can be found in Fig. 6 of 
Ref.~\cite{Pattabi-PLB-163Tm-07}. The occupation of the $i_{13/2}$ proton 
level in the Lu isotopes adds an additional degree of shape driving towards 
larger deformation. However, as stated above, the data indicate that its 
impact is rather minor. This is borne out by the calculations where average 
deformations of ${\epsilon}_2=0.39$, $|\gamma|=17^{\circ}$ for $^{163}$Tm should 
be compared with computed values of ${\epsilon}_2=0.41$, ${\gamma}=+19^{\circ}$ 
for $^{163}$Lu. The nearly equal deformations find their origin in the 
following physical effect: $^{163}$Lu does not make full use of the $N=94$ 
gap, because it has two fewer neutrons, but this absence is compensated by 
the additional drive provided by the $i_{13/2}$ proton. As argued in 
Ref.~\cite{Pattabi-PLB-163Tm-07}, the large $N=94$ gap makes it unlikely 
that the $^{163}$Tm TSD bands involve a three-quasiparticle structure with 
a proton coupled to a neutron particle-hole excitation. 
The possibility that these bands correspond to configurations with the 
odd proton occupying the $[541]1/2^{-}$ level (labeled as $h_{9/2}$ in 
Fig. 6 of Ref.~\cite{Pattabi-PLB-163Tm-07}) is also unlikely. This 
orbital is characterized by a large signature splitting and small 
$B(M1)$ values for inter-band transitions, in clear contradiction 
with the data~\cite{Pattabi-PLB-163Tm-07}. While it is possible that 
combining the occupation of the $[541]1/2^{-}$ level with a neutron particle-hole 
excitation would alter these observables, it would result in an excitation 
energy much larger than seen experimentally because of the $N=94$ gap. 
Furthermore, as can be seen in Fig. 6 of Ref.~\cite{Pattabi-PLB-163Tm-07}, 
there are no other low-lying proton excitations that lead to small signature splitting. 

Finally, it is worth noting that the values of the $Q_{sf}$ moments associated 
with the sidefeeding differ significantly between bands 1 and 2, $Q_{sf}$ $\sim$ 
$6.8~eb$, and bands TSD1 and TSD2, $Q_{sf}$ $\sim$ $10~eb$ 
(see Table~\ref{tab:fitmoments}). This change in $Q_{sf}$ values is in part 
responsible for the large difference in the $F(\tau)$ curves as a function of 
energy seen in Fig.~\ref{fig:curve}. It implies that the $\gamma$-ray intensity 
responsible for the feeding of the bands originates from states associated with 
different intrinsic structures. The calculations presented in Ref.~\cite{Pattabi-PLB-163Tm-07} 
suggested that several other TSD bands, corresponding to various ph excitations, 
should be present in $^{163}$Tm at excitation energies comparable 
to those of bands TSD1 and TSD2. It is plausible that these other TSD bands are 
part of the final stages in the deexcitation process towards the yrast TSD 
bands. If this is the case, the present observations suggest 
that the average deformation associated with the feeding TSD bands is larger 
than that of their yrast counterparts. Conversely, the feeding of bands 1 and 
2 then appears to occur from levels associated with a smaller deformation, 
similar to that of the bands themselves. 

\section{\label{sec:summary}Summary and Conclusion}

The transition quadrupole moments, $Q_t$, of two recently observed 
TSD bands in $^{163}$Tm have been determined with the 
DSAM method and compared with the moments measured 
for the yrast signature partner bands of normal deformation. While the 
data confirm that the TSD bands are associated with a larger deformation, 
the measured $Q_t$ moments are smaller than the calculated values. It 
was pointed out that this difference between theory and experiment appears 
to a general feature of the region, which requires further investigation. 
The data also indicate that the feeding of the TSD bands is associated 
with states of larger deformation.

\section*{ACKNOWLEDGMENTS}
The authors thank J. P.~Greene for the preparation of the target, C. Vaman, 
D. Peterson, and J. Kaiser for assistance with some of the computer codes 
used in the present work and D. J.~Hartley for fruitful discussions. 

This work has been supported in part by the U.S. Department of Energy, Office 
of Nuclear Physics, under contract No. DE-AC02-06CH11357, the U.S. 
National Science Foundation under grants No. PHY04-57120 and INT-0111536, and 
the Department of Science and Technology, Government of India, under grant 
No. DST-NSF/RPO-017/98.

\end{document}